\documentclass[aps,pra,twocolumn,a4paper]{revtex4-1}

\usepackage{amsmath,amsfonts,amssymb,amsthm}
\usepackage{eqnarray}
\usepackage{mathcomp} 
\usepackage{srcltx}
\usepackage{ulem} 

\usepackage{graphics,graphicx}
\usepackage{dcolumn}
\usepackage{bm}
\usepackage{hyperref}
\usepackage[mathlines]{lineno}
\usepackage{siunitx} 
\usepackage[dvipsnames]{xcolor} 
\usepackage{units} 
\usepackage{color,colortbl}

\usepackage{ragged2e}
\usepackage{booktabs} 
\usepackage{multirow,tabularx}

\usepackage{fancyhdr}
\usepackage{array}
\newcolumntype{L}[1]{>{\raggedright\let\newline\\\arraybackslash\hspace{0pt}}m{#1}}
\newcolumntype{C}[1]{>{\centering\let\newline\\\arraybackslash\hspace{0pt}}m{#1}}
\newcolumntype{R}[1]{>{\raggedleft\let\newline\\\arraybackslash\hspace{0pt}}m{#1}}

\newcommand{\ket} [1] {|#1\rangle}
\newcommand{\beq}{\begin{equation}}
\newcommand{\eeq}{\end{equation}}
\newcommand{\bea}{\begin{eqnarray}}
\newcommand{\eea}{\end{eqnarray}}

\def\({\left(}
\def\){\right)}
\def\[{\left[}
\def\]{\right]}

\definecolor{LightGray}{gray}{0.9}

\begin{document}

\title{Experimental quantum key distribution beyond the repeaterless secret key capacity}

\author{M. Minder$^{1,2 \dagger}$, M. Pittaluga$^{1,3 \dagger}$, G. L. Roberts$^{1,2}$, M. Lucamarini$^{1\star}$, J. F. Dynes$^{1}$, Z. L. Yuan$^{1}$ \& A. J. Shields$^{1}$ \endgraf
\small\itshape
$^{1}$Toshiba Research Europe Ltd, 208 Cambridge Science Park, Cambridge CB4 0GZ, UK \\
$^{2}$Department of Engineering, Cambridge University, 9JJ Thomson Avenue, Cambridge CB3 0FA, UK \\
$^{3}$School of Electronic and Electrical Engineering, University of Leeds, Leeds LS2 9JT, UK \\
\smallskip
$^{\dagger}$These authors contributed equally to this work.\\
$^{\star}$Email to: marco.lucamarini@crl.toshiba.co.uk\\
}

\begin{abstract}
Quantum communications promise to revolutionise the way information is exchanged and protected.
Unlike their classical counterpart, they are based on dim optical pulses that cannot be amplified by conventional optical repeaters.
Consequently they are heavily impaired by propagation channel losses, which confine their transmission rate and range below a theoretical limit known as repeaterless secret key capacity.
Overcoming this limit with today's technology was believed to be impossible until the recent proposal of a scheme that uses phase-coherent optical signals and an auxiliary measuring station to distribute quantum information.
Here we experimentally demonstrate such a scheme for the first time and over significant channel losses, in excess of 90~dB.
In the high loss regime, the resulting secure key rate exceeds the repeaterless secret key capacity, a result never achieved before.
This represents a major step in promoting quantum communications as a dependable resource in today's world.
\end{abstract}

\maketitle

Given the colossal amount of digital information transmitted daily and the stringent security requirements often needed, it is no wonder that quantum communication is attracting increasing attention from the scientific community and of the general public. In particular, quantum key distribution (QKD)~\cite{BB84,Ekert.1991} enables secure communication between two parties (Alice and Bob) without making assumptions on the computational power of a potential eavesdropper (Eve). The intriguing capabilities of QKD have motivated intense research to readying it for real world and everyday use. On one hand, research focusses on the seamless integration of QKD into existing optical networks~\cite{Tow97,Frohlich.2013}, as this would enable the widespread use of QKD through a trusted-node architecture~\cite{PPA+09}.
On the other hand, when the highest level of security is demanded, it becomes crucial to avoid, at least partially, the trusted intermediate nodes, which have no quantum protection against Eve. In this respect, a practical solution would be measurement-device-independent (MDI) QKD~\cite{Lo.2012} (see also~\cite{BP12,Tamaki.2012}), which allows two parties to communicate via an intermediate, completely untrusted, node.
A complete solution would be quantum repeaters~\cite{BDCZ98,DLCZ01,SSdRG11,GKF+15}, but they are still outside today's technological reach.

All forms of QKD, however, are only possible under a certain amount of loss. If $1-\eta$ is the total loss of the quantum channel, the secure key rate of QKD scales at best linearly with $\eta$~\cite{TGW14,PLOB17}.
This is a fundamental physical limit, known as ``repeaterless secret key capacity'' (SKC$_0$)~\cite{PLOB17}.
The subscript `0' emphasises the absence of any repeater between the users, even a simple trusted node that distils one key with Alice and one with Bob. So the SKC$_0$ quantifies the maximum secret information that can be exchanged on a direct optical link.

In MDI-QKD, as well as in entanglement-based QKD~\cite{Ekert.1991,BBM92}, there is an intermediate untrusted node, usually called Charlie, that relays the quantum signals between the users.
Therefore, in principle, these schemes could achieve a better capacity 
than QKD and even overcome the SKC$_0$~\cite{Pir16} using only untrusted nodes.
This possibility has been described theoretically for adaptive MDI-QKD~\cite{ATM15} as well as for entanglement-based QKD~\cite{LJKL16} and MDI-QKD, when both are assisted by quantum memories~\cite{AKB14,PRML14}.
However, despite being simpler than a full-fledged quantum repeater~\cite{BDCZ98,DLCZ01,SSdRG11,GKF+15}, these schemes still rely on quantum memories, which are not practical. As a result, no experiment has yet overcome the SKC$_0$ using only intermediate untrusted nodes.

Recently, it was theoretically shown that a phase-based setup realisable with existing, standard, off-the-shelf components has the potential to overcome the repeaterless bound~\cite{LYDS18}.
The new scheme, named twin-field QKD (TF-QKD) for its use of remotely prepared optical fields with similar (i.e. twin) phases, provides a key rate that scales with $\eta^{1/2}$, overcomes the SKC$_0$ at long distances and, like MDI-QKD, makes use of an intermediate node that can be completely untrusted.
TF-QKD was initially proven secure under restrictive assumptions on Eve~\cite{LYDS18} which have recently been removed~\cite{Tamaki.,Ma.,Wang.2018,CYW+18,CAL18,LL18}, thus providing unconditionally secure TF-QKD protocols.

In TF-QKD, Alice and Bob encode their information in two optical pulses that are sent to a central beam splitter, located in the intermediate station. Random phases are initially added to the pulses, or to part of them, to improve the performance of the scheme through the use of decoy states~\cite{Hwang.2003,Wan05,LMC05}.
The information retrieved from the detection events, announced by Charlie, is used to form a secret key, or to test the channel against the presence of Eve, or both.
Many variants of this basic scheme are possible. The most practical ones, based on coherent states, are characterised by some crucial ingredients:
the first-order interference of the optical pulses at the intermediate node and their subsequent single-photon detection;
the phase randomisation of some of the pulses followed, in some cases, by the reconciliation of the phases in post-processing.
To implement these features, the distant users need to phase-stabilise the optical paths connecting them to the intermediate node.
Even with perfectly prepared pulses, the microscopic length variation in either path is sufficient to prevent the high-visibility interference required for successful TF-QKD.

Here we report the first experiment where a positive key rate is obtained using TF-QKD.
The resulting key rate overcomes the SKC$_0$ limit at high channel losses and remains positive up to about 90~dB.
This is 100~times more loss than in previous record-distance experiments with QKD~\cite{BBR+18} and MDI-QKD~\cite{YCY+16} and a total loss five orders of magnitudes larger than in satellite QKD~\cite{LCL+17}.
It represents the first experimental evidence that the repeaterless secret key capacity bound can be overcome.

\section*{Experimental setup}

\begin{figure*}
    \centering
    \includegraphics[width=0.85\textwidth]{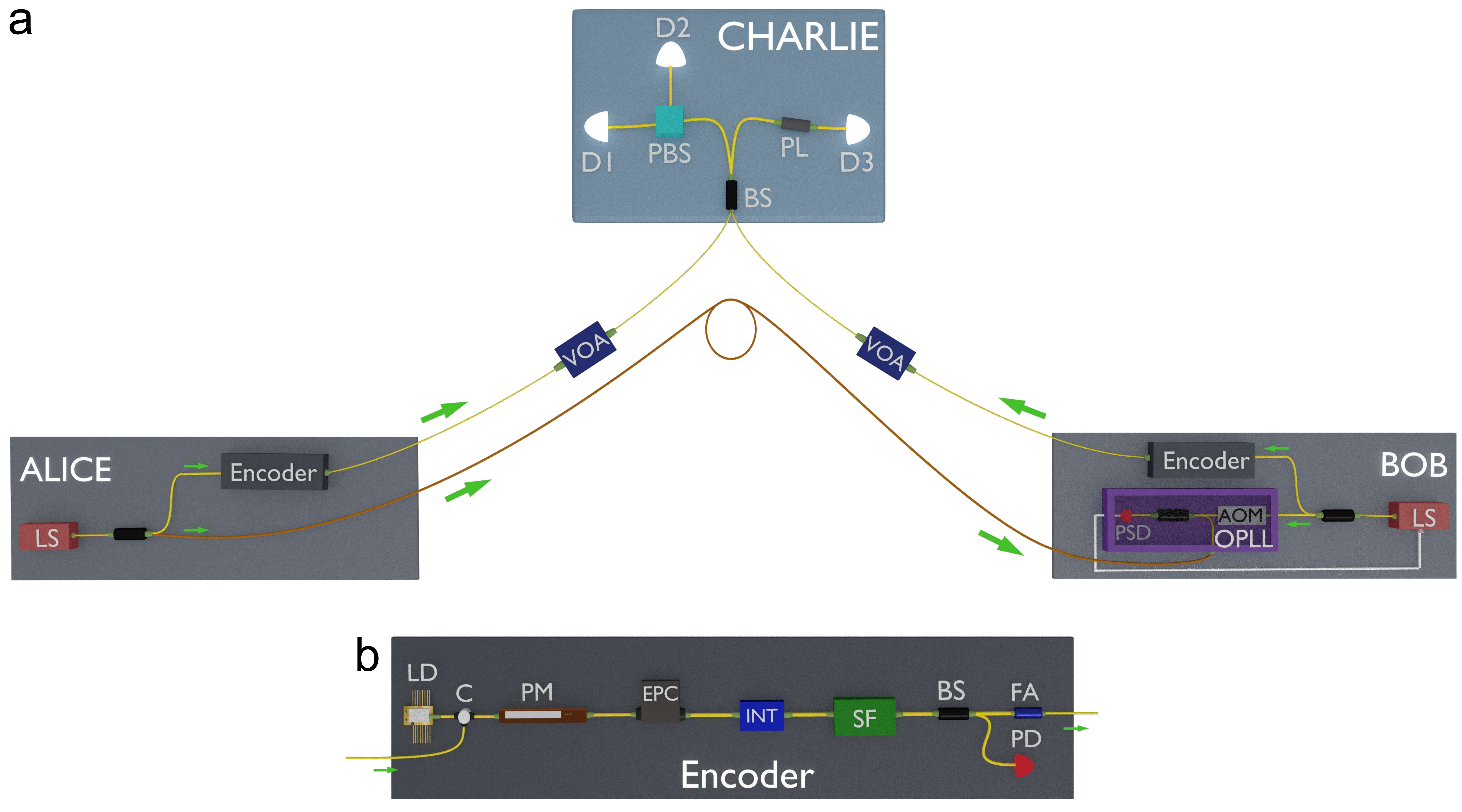}
    \caption{
    \textbf{Experimental setup. a,}
    Alice and Bob generate light beams from their local continuous-wave laser sources (LSs) and send them onto their beam splitters.
    From each beam splitter, one output beam goes to the Encoders and is used to perform TF-QKD.
    The other is used to lock the users' LSs through a service fibre, depicted in orange.
    Alice sends part of her light to Bob through the service fibre.
    Bob interferes it with his own, after shifting its frequency by 80~MHz through an acousto-optic modulator (AOM).
    The beating signal resulting from the interference is detected with a phase-sensitive detector (PSD) whose current is proportional to the phase difference between Alice's and Bob's light beams.
    An electronic feedback is then given to Bob's LS based on the detected difference to lock its phase to Alice's one.
    This constitutes a heterodyne optical phase-locked loop (OPLL).
    \textbf{b,}
    In the Encoder modules, the continuous-wave light prepared locally by Alice's and Bob's LSs, seeds a gain-switched laser diode (LD) that carves it into pulses.
    The optical pulses are either rapidly modulated or finely controlled in phase by the phase modulators (PMs).
    After crossing the electrical polarisation controller (EPC) and the intensity controller (INT), part of the pulses is directed to the power detector (PD) for monitoring the intensity and the other part travels through the quantum channel towards Charlie's beam splitter (BS).
    Here, they interfere with the other user's pulses.
    The outcome of the interference is registered by the SNSPD detectors D1, D2 and D3.
    Variable optical attenuators (VOAs) add losses to the quantum channel.
    C, circulator; SF, spectral filter; FA, fixed attenuator; PL, polariser; PBS, polarising beam splitter. 
    }
    \label{fig:setup}
\end{figure*}

In our realisation of TF-QKD, we consider a generalised protocol (see Methods) that can be modified to encompass various TF-QKD protocols based on coherent states~\cite{LYDS18,Tamaki.,Ma.,Wang.2018,CYW+18,CAL18}.
To experimentally validate these protocols and overcome the SKC$_0$ bound, Alice and Bob should use two separate lasers to prepare coherent states in a given phase and polarisation state, with various intensities.
The two separate lasers should be phase-locked to let the users reconcile their phase values.
We represent Alice's states as $\ket{\sqrt{\mu_a} e^{i\varphi_a}}$, where $\mu_a$ is the intensity and $\varphi_a \in [0,2\pi)$ is the phase.
Bob prepares similar states with the subscript $a$ replaced by $b$.
The phases $\varphi_{a,b}$ include both the bit information and the random values needed in coherent-state TF-QKD.
The optical pulses emitted by the users should interfere with high visibility in the intermediate station after having travelled through a pair of highly lossy channels.
High loss is needed to overcome the SKC$_0$~\cite{LYDS18}.
The optical phase should remain stable in time, which is challenging when the channel loss reduces the amount of detected counts.

We implement these features using the experimental setup shown in Fig.~1. 
Each user is endowed with a continuous-wave laser source (LS).
Alice's LS acts as the phase reference.
Its light is split in two at a first beam splitter (BS).
One part is sent to Bob through a service fibre, depicted in orange in Fig.~1
a, and is used to lock Bob's LS via a heterodyne optical phase-locked loop (OPLL~\cite{Bordonalli.1999}, see Supplementary Section~1 for further details).
In Bob's module, the reference light interferes with another light beam prepared by Bob and shifted by 80~MHz by an acousto optic modulator (AOM).
A photodiode acts as a phase-sensitive detector (PSD), whose intensity is mapped onto a phase difference $\delta\varphi$ between the reference light and Bob's local light.
A feedback is then given to Bob's laser based on the value of $\delta\varphi$.
With this OPLL, Bob's laser is locked to Alice's with a phase error less than $5^{\circ}$, which includes a potential phase fluctuation in the fibre connecting Alice to Bob.
An attacker could modify the reference light while it travels from Alice to Bob, but that would not affect the security of the scheme.
Any modification would translate into a different value of $\delta\varphi$, which is equivalent to Eve introducing phase noise on the main channels going from the users to Charlie (see also~\cite{Koa04} for a similar argument applied to QKD).
However, we do not claim here the robustness of Alice's and Bob's modules to side-channel attacks, which requires more scrutiny, similarly to the one ongoing for the MDI-QKD sending modules.

The fraction of each user's light not involved in the phase-locking mechanism is directed to the Encoder, depicted in Fig.~1b. 
Here it enters the cavity of a slave laser diode (LD) that is periodically gain-switched to produce a pulse train at 2~GHz.
This ensures that each pulse will inherit the phase of the injected optical field, which is locked to the reference light.
Moreover, Alice and Bob's LDs will emit pulses as narrow as \SI{70}{ps} at 1548.92~nm, with high extinction ratio and constant intensity due to the strength of optical injection into the slave laser being 1,400 times weaker than the electrical injection, as we measured. After the LD, the optical pulses pass through an in-line phase modulator (PM), which applies fast modulation from an RF signal to encode the phase values required by the specific TF-QKD protocol, and a slow correction from a DC signal to compensate the phase noise on the paths linking to Charlie.
After setting the optical pulses' polarisation and intensity, the pulses pass through 15~GHz filters that clean their spectral mode~\cite{CLF+16a}, thus ensuring high visibility interference between the twin-fields. Then they are sent to variable optical attenuators (VOAs) that vary the loss of the channel connecting the users to Charlie.

Alice and Bob's optical fields interfere on Charlie's BS and are eventually detected by superconducting nanowire single photon detectors (SNSPDs, Single Quantum EOS 410 CS) cooled at 3.2~K, featuring 22~Hz dark count rate and 44\% detection efficiency.
Detector D1 is associated with a 100~ps resolution time tagger and is used to extract the raw key rate.
D2 monitors the optical field leakage into the non-intended polarisation, which is minimised by Alice and Bob through their polarisation controllers.
D3 is sampled by a photon counter at a minimum interval of 10~ms to stabilise the overall phase.

\section*{Results}

The first task is providing the users with weak coherent pulses that are locked to a common phase reference and capable of interfering on Charlie's BS. Then part of these pulses have to be phase randomised with respect to the phase reference. In some TF-QKD protocols~\cite{LYDS18,Tamaki.,Ma.,Wang.2018}, it is necessary that the users know the values of the random phases, whereas in others~\cite{CYW+18,CAL18,LL18} this is not mandatory.

In the current setup we randomise the phase in an active way and obtain a first-order interference visibility at Charlie of 96.4\% when the OPLL is active and the two PMs in Alice and Bob encode equal phases.
We encode a pseudo-random pattern containing $2^{10}$ symbols having $2^5$ modulation levels through the PMs driven by high-speed 12~GSa/s digital-to-analogue converters (DACs) with 8-bit amplitude resolution.
The number of phases we chose is sufficiently close to a phase randomisation with infinite random phases~\cite{CZLM15}.
However, to further demonstrate a full phase randomisation, we performed a parallel experiment using a continuous phase randomisation from a master gain-switched laser~\cite{JCS+11,YLD+14} (see Supplementary Section~2).
For that, we removed the OPLL and active phase randomisation from the setup, while leaving active phase encoding.
Then we locked Alice and Bob's lasers to the main master laser by optical injection locking~\cite{YPJ+03,CLF+16a} and obtained a visibility of 97.5\%, 1.1\% higher than in the previous case.
We attribute this difference to the absence of errors from the OPLL and the active phase randomisation.
This result shows that the overall visibility is not affected by the absolute number of encoded phases (it is higher with more phases than less) as much as by the components used to implement it.
The base quantum bit error rate (QBER) of the system remains in all cases smaller than 1.8\% and there is no in-principle limitation to increasing the number of the encoded phases.

The ambient temperature fluctuations cause the experimentally obtained interference to drift.
Even with 40~metres of optical fibre, the environmental fluctuations cause a relative phase drift in our setup of \SI{0.7}{rad/s}.
This requires a feedback control every 10 -- 100~ms to avoid detrimental effects on the QBER.
To implement this phase control, we doubled the pulse pattern to $2048$ bits and temporally interleaved phase-encoded pulses and unmodulated reference pulses, with equal duty cycles and intensities.
This is done by clocking the phase modulators in Alice and Bob's setup at 2~GHz and by actively switching between reference pulses and encoding pulses.
This reduces the effective clock rate of the TF-QKD protocol to 1~GHz.
The photon detection clicks are recorded by Charlie with time-tagging electronics and are grouped in post-processing according to their phase values.
From these events we retrieve the gain and the QBER of the system (Supplementary Section~4).
The phase correction from reference pulses is designed to keep Alice and Bob's optical fields locked on Charlie's BS at a constant $\pi/2$ phase difference.
This is the most efficient solution as it exploits the linear part of the response function (see Supplementary Section~3 for further discussion).
The phase offset is continuously monitored by detector D3 and corrected by acting on the DC level of one of the PMs in the transmitting modules.
This is equivalent to having an extra PM in Charlie's station, as proposed in ref.~\cite{LYDS18}.

A main advantage of TF-QKD is the scaling property of the secret key rate with the square-root of the channel transmission, $\eta^{1/2}$.
This would be impossible without correspondingly having the square-root scaling of the detection rate.
We verified this essential feature of TF-QKD directly and summarised the result in Fig.~2.
\begin{figure}
    \centering
    \includegraphics[width=\columnwidth]{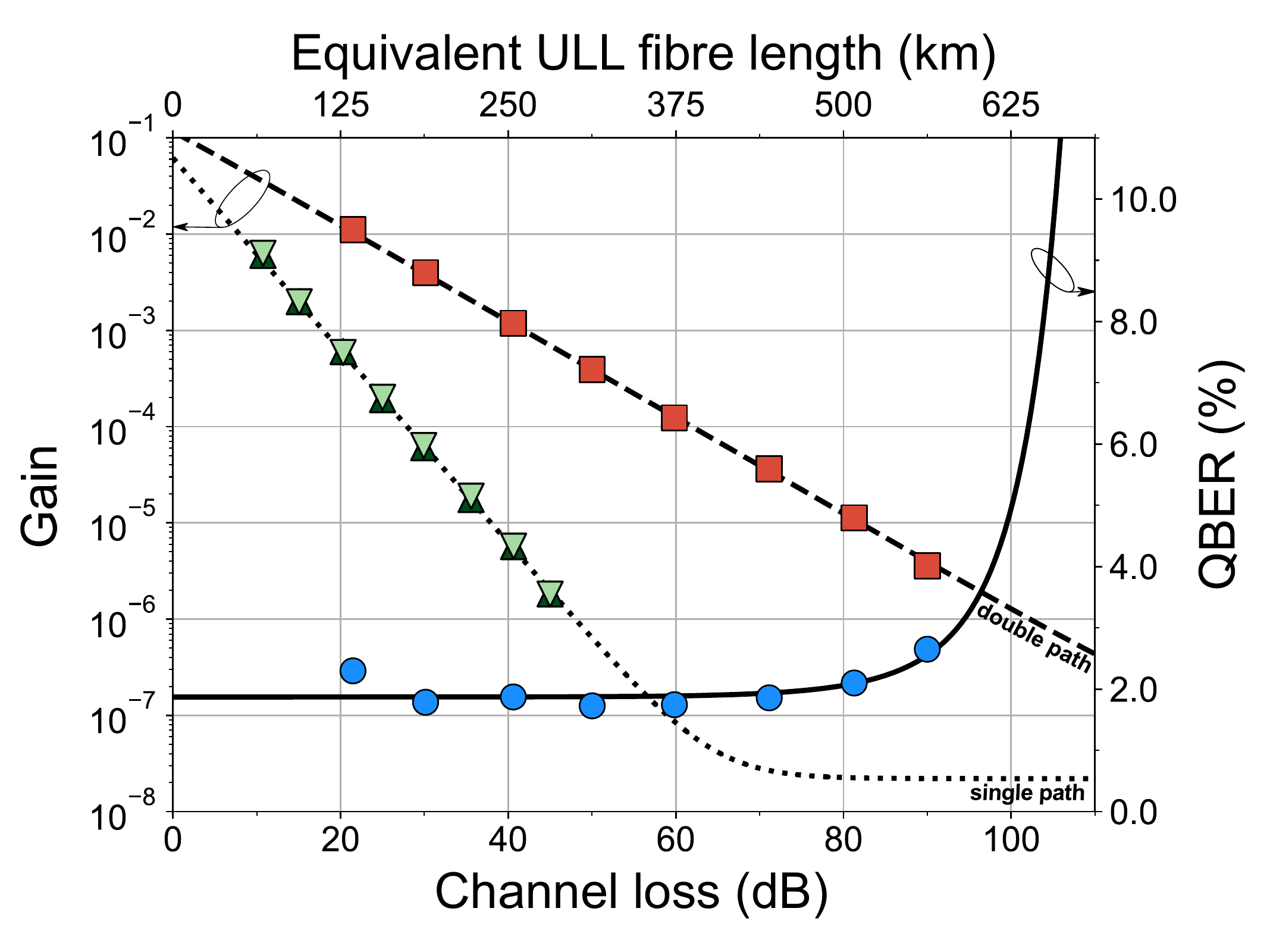}
    \caption{
    \textbf{Gain and QBER of TF-QKD.}
    The gain is the detection probability per encoding gate.
    Gain and QBER are plotted against the channel loss $1-\eta$.
    The equivalent fibre length on the top axis pertains to an ultra-low-loss (ULL) fibre with attenuation coefficient 0.16~dB/km.
    The different scaling laws of the gain for a QKD-like single-path quantum transmission and for a TF-QKD-like double-path quantum transmission are apparent.
    The upward and downward triangular (square) points on the dotted (dashed) line are the single-path (double-path) experimental detection rate recorded for different channel loss.
    The circle points on the solid line are the double-path experimental QBER of the interfering pulses.
    All the experimental data agrees well with the theoretical curves.
    }
    \label{fig:gain_TFQKDvsQKD}
\end{figure}
The data corresponding to a direct-link quantum transmission were taken by shutting off one arm of our experimental setup, thus allowing a single user at a time to signal to Charlie's station.
The data for double-path transmission, on the other hand, were taken with both arms open.
As is apparent from the figure, the single-path gain (triangular points on the dotted line) scales linearly with the loss, $1-\eta$, whereas the double-path gain (square points on the dashed line) scales with the square-root.
At any given gain, the double-path TF-QKD can tolerate channel loss twice as large than single-path direct-link QKD.
In the same figure, we also report the experimental QBER of TF-QKD, which is composed of three main contributions: the quantum state preparation, detectors' dark counts and feedback routine.
In our setup, the last two terms significantly affect the overall QBER only at losses higher than 70~dB.

Our experimental results are independent of the specific security analysis adopted to extract a key rate.
Hence they can be used as a reference to test the performance of any TF-QKD-like protocol.
Here we analyse the data for three TF-QKD protocols, two~\cite{LYDS18,Wang.2018} over the whole loss range and one~\cite{CAL18} at a specific loss around 70~dB.
The protocol in ref.~\cite{LYDS18} is the original TF-QKD scheme and acts as a reference.
The protocols in refs.~\cite{Wang.2018,CAL18}, on the other hand, have been conceived to be unconditionally secure.
We have implemented them using three intensities, which is practical as compared with infinite intensities in their initial proposals.

\begin{figure*}
    \centering
    \includegraphics[width=0.8\textwidth]{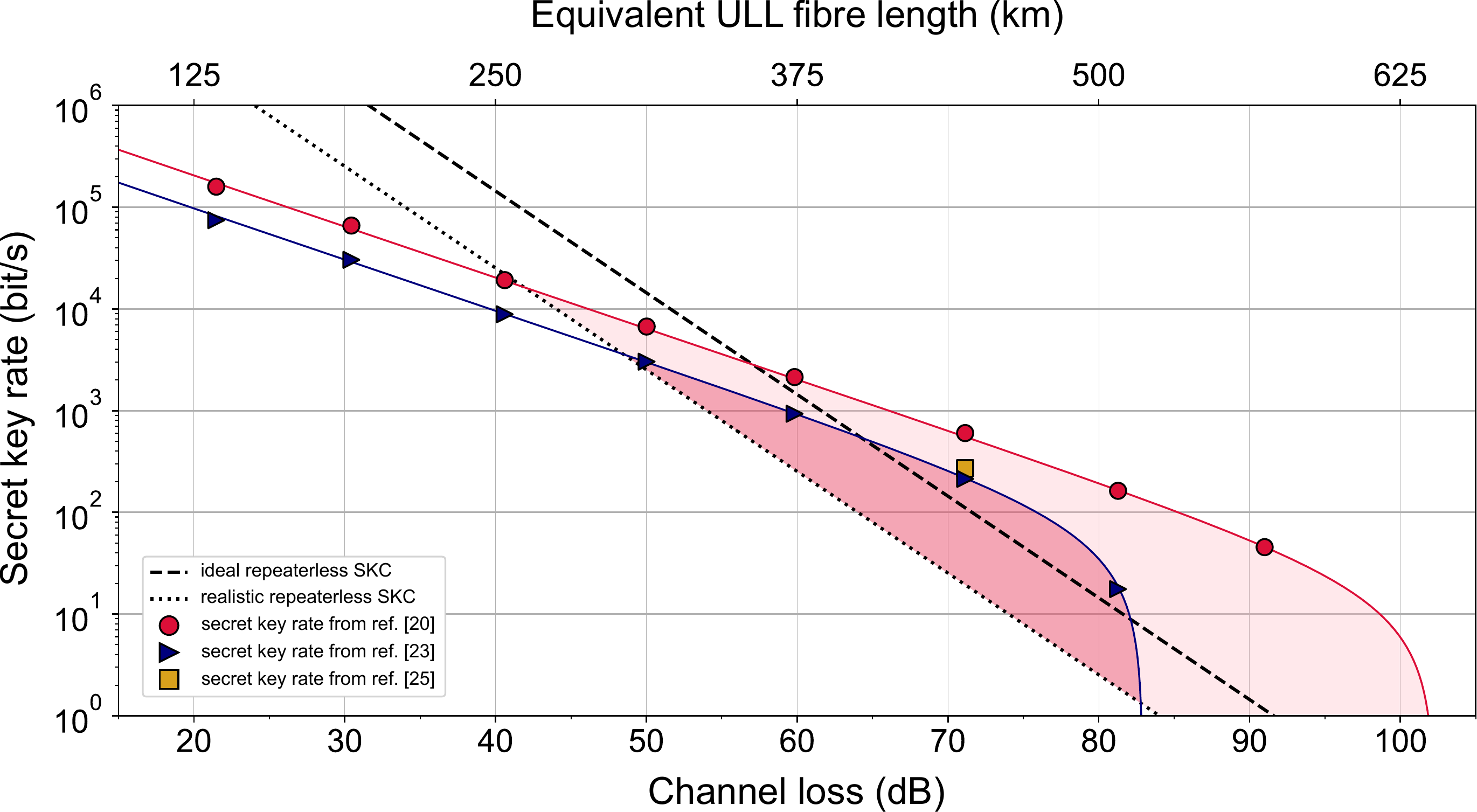}
    \caption{
    \textbf{TF-QKD key rates.}
    Secret key rates are plotted against the channel loss (lower horizontal axis) and the corresponding ULL fibre distance (upper horizontal axis).
    The markers show the acquired experimental data whereas the solid lines follow from simulations.
    The ideal repeaterless SKC$_0$~\cite{PLOB17} (dashed line) and the realistic one (dotted line) are plotted along with the key rates of the original TF-QKD protocol~\cite{LYDS18} (red circles), the protocol in ref.~\cite{Wang.2018} (blue triangles) and the protocol in ref.~\cite{CAL18} (yellow square).
    The TF-QKD supremacy region is in pink shades.
    The simulations assume 1~GHz effective clock rate.
    The realistic repeaterless bound assumes a total detection efficiency of 35\% plus 3~dB loss due to having one detector in Charlie's module.
    Other parameters are:
    $\alpha=0.16$~dB/km, ULL fibre attenuation;
    $f_{EC}=1.15$, error correction coefficient.
    $\eta_{C} = 30.8\%$, total transmission of Charlie's module, resulting from $\eta_{det}=44\%$ and $\eta_{\textrm{coupling}}=70\%$;
    $P_{dc} = 22$~Hz, dark count rate.
    Charlie is assumed to be at equal distance from Alice and Bob.
    The photon fluxes are specified in the Methods.
    }
    \label{fig:SKR}
\end{figure*}

In Fig.~3 
we plot the secure key rates (SKRs) versus the channel loss for the protocols analysed.
We also plot two lines for the repeaterless secret key capacity, which we call \textit{ideal} and \textit{realistic} SKC$_0$, respectively.
The former is the expression given in~\cite{PLOB17}, $\log_{2}[1/(1-\eta)]$, with detectors implicitly assumed to be 100\% efficient.
This SKC$_0$ is impossible to overcome without an intermediate repeater.
The latter represents a direct comparison with our experiment and assumes a QKD performed with one detector and efficiency slightly larger than in our setup.

The darker-pink (lighter-pink) shaded area is the supremacy region where the SKR of the protocol in~\cite{Wang.2018} (the protocol in~\cite{LYDS18}) surpasses the realistic SKC$_0$.
This region extends from about 50~dB to 83~dB, limited only by detectors' dark counts.
In this range, our TF-QKD scheme provides more SKR than a QKD scheme with the same components.
This is remarkable in light of the fact that TF-QKD is more secure than QKD, as it protects against attacks directed at the detection devices.
Even more interestingly, there are experimental points that fall beyond the \textit{ideal} SKC$_0$.
At 71.1~dB, for instance, the SKRs of the protocols in refs.~\cite{Wang.2018} and \cite{CAL18} are $213.0$~bit/s and $270.7$~bit/s, respectively, i.e.~1.90 times and 2.42 times larger than the corresponding ideal SKC$_0$ ($112.0$~bit/s).
This is the first time that this fundamental limit has been overcome experimentally.
It is worth mentioning that all the reported SKRs are quite conservative as they include the penalty due to an imperfect error-correction ($f_{EC}=1.15$).
The maximum channel loss over which we can stabilise the phase and obtain a positive key rate is 90.8~dB (rightmost red circle in Fig.~3
).
This is equivalent to 454~km and 567~km of standard and ultralow-loss (ULL, 0.16~dB/km) single-mode optical fibre, respectively, connecting the users.
We incidentally notice that values up to 0.1419~dB/km are currently achievable in fibres at 1560~nm~\cite{TSM+18}.

It is interesting to compare these results with the current record distances~\cite{BBR+18,YCY+16} obtained in QKD and MDI-QKD experiments over long-haul fibres in the finite-size scenario.
The QKD record distance is 421.1~km in ULL fibre, with a key rate of 0.25~bit/s over a total channel loss of 71.9~dB~\cite{BBR+18}.
This was made possible by a 2.5~GHz clock rate and a detector dark count rate of 0.1~Hz.
For a similar channel loss, with our 1~GHz effective clock rate, the SKR of any of the TF-QKD protocols analysed is three orders of magnitude higher.
The longest demonstration of MDI-QKD is on 404~km of ULL optical fibre obtained with the protocol in~\cite{ZYW16}.
With a clock rate of 75~MHz, it provided an SKR of \num{3.2e-4}~bit/s over a total channel loss of 64.64~dB~\cite{YCY+16}, which is six orders of magnitude smaller than the TF-QKD key rates at 71.1~dB.
Although our results have been achieved in the asymptotic regime and do not include finite-size effects or long-haul real fibres, as the experiments in~\cite{YCY+16,BBR+18}, the improvement they entail appears to be substantial.

\section*{Conclusions}

In TF-QKD, quantum information is carried by the optical fields prepared by Alice and Bob.
Fields can tolerate larger loss than photons and can potentially increase the rate and range of quantum communications~\cite{LYDS18}.
In the present work, we have provided the first experimental evidence of this potential, surpassing, for the first time, the rate-loss limit of direct-link quantum communications~\cite{PLOB17} with an intermediate untrusted node.

To achieve this goal, the users have to prepare light pulses that are phase stabilised and at the same time phase randomised respect to a shared phase reference. The phase stabilisation between the lasers was achieved by an optical phase-locked loop, which also guarantees that the reference light emitted by Alice and possibly manipulated by Eve is securely transferred to Bob.
The phase stabilisation across the channels linking Alice and Bob to Charlie was achieved using efficient feedback based on unmodulated reference optical pulses of the same intensity as the quantum pulses used for key generation.
This maintains interference stability to overcome the rate-loss theoretical limit and extract a positive key rate over a 90~dB loss link, which is about 20~dB larger than in any other previous quantum communication test.

Our proof-of-concept experiment shows that TF-QKD can greatly enhance the range and rate of quantum communications using presently available technology.

\section*{Methods}
\smallskip

{\small
\noindent\textbf{Generalised TF-QKD protocol}
\smallskip

\textit{State Preparation} -- Alice randomly selects: the bit value $\alpha_a=\{0,1\}$, with probability $p_{\alpha_a}$; the basis value $\beta_a=\{0,1\}=\{Z,X\}$, with probability $p_{\beta_a}$; the global phase value $\phi_a\in [0,2\pi)$, with uniform probability $p_{\phi_a}$; the intensity value $\mu_a=\{u_a,v_a,w_a\}$, with probability $p_{\mu_a}$. She uses the setup in Fig.~1
to prepare a coherent state $\ket{\sqrt{\mu_a}e^{i\varphi_a}}$, where $\varphi_a=\phi_a+\alpha_a\pi+\beta_a\pi/2$. Bob does the same, with subscripts $a$ replaced by $b$ and same values for the parameters, unless explicitly stated.

This step represents the state preparation of the original TF-QKD protocol~\cite{LYDS18} and of the one in ref.~\cite{Tamaki.}.
To increase the asymptotic key rate, the probability of the majority basis, $p_Z$, can be set arbitrarily close to 1, similarly to the efficient version of the BB84 protocol~\cite{LCA05}.
Along the same line, the state preparation of the TF-QKD protocol in ref.~\cite{Ma.} can be obtained by simply setting the probability of the minority basis, $p_{X}$, equal to 0.
In ref.~\cite{CYW+18} and in the Protocol~3 of ref.~\cite{CAL18}, which relies on coherent states, the global phase is randomised only in the `test' basis, which we choose here equal to $X$, while it is constant in the `encoding' basis, which we choose here equal to $Z$.
For the state preparation of the protocol in ref.~\cite{Wang.2018}, we can treat the $X$ basis as in the original TF-QKD protocol whereas for the $Z$ basis we set $w_{a,b}=0$ and $p_{u_{a,b}}=1-p_{w_{a,b}}=\epsilon$, $p_{v_{a,b}}=0$.
A close-to-optimal value for $\epsilon$ is $10\%$. In our simulations, we set it equal to 7.8\%. We then relate the bit value $0$ ($1$) to the instances of the $Z$ basis where Alice encoded $w_a$ ($u_a$) and Bob encoded $u_b$ ($w_b$).

In the above preparation of the intensity, we considered for simplicity only three values, as opposed to the infinite values considered in most of the TF-QKD proposals.
On one hand, this setting can easily be generalised to any number of intensities, even infinite, like in the original decoy-state QKD~\cite{LMC05}.
On the other hand, the possibility to import the decoy-state technique into TF-QKD is the main motivation for having multiple intensities in the protocol.
The only exception is ref.~\cite{LL18}, which we could not include in the above description as it does not resort to phase randomisation and decoy states to extrapolate a key rate from the acquired sample.
For the other protocols, we describe below how to apply the decoy-state technique with three intensities.

\smallskip

\textit{Measurement} -- Alice and Bob send their optical pulses to the central relay station, Charlie, who does not need to be honest.
A honest Charlie would send the incoming pulses on his beam splitter, measure the output pulses and report which of his two detectors clicked.
A dishonest Charlie, however, could use any detection scheme he pleases.
This would not affect the security of the protocol as TF-QKD's security, similarly to MDI-QKD, is independent of the detection scheme.
This implies that the secret key rate extracted by the users when Charlie is dishonest is always lower than or equal to the one they would extract if Charlie is honest.
In our experiment, Charlie announces counts from the detector D1 only.

\smallskip

\textit{Announcement} -- After repeating the above steps many times, the honest Charlie announces over a public authenticated channel the events where one and only one detector clicked. Alice and Bob announce their intensity values $\mu_{a,b}$, their basis values $\beta_{a,b}$ and their global phases $\phi_{a,b}$.

This step holds for protocols in~\cite{LYDS18,Ma.}, even if the announcement of the basis is redundant in~\cite{Ma.} because $p_X=0$. For the protocols in~\cite{Wang.2018,CYW+18,CAL18}, the intensity values are announced only in the $X$ basis, whereas the global phases are announced only in the $X$ basis for \cite{Wang.2018} and never announced in \cite{CYW+18,CAL18}. This latter feature is remarkable and can entail a great experimental simplification. For the protocol in \cite{Tamaki.}, Alice selects two modes of execution. In the `test mode' the global phases are never disclosed, thus allowing a rigorous application of the decoy-state method; in the `code mode', the global phases are announced to let the users reconcile their bit values. The bases are always announced unless a particular event occurs in the code mode.
\smallskip

\textit{Sifting} -- Among the announced successful detections, Alice and Bob keep the events that have matching values.
In all cases, either for a specific basis or for both bases, they keep those events whose phases are `twins', i.e., no more different than a certain tolerance level $\Delta$  modulo $\pi$, due to the symmetry of TF-QKD with respect to the addition of $\pi$ to the phase values~\cite{Ma.}.
After sifting their data, the users keep $\alpha_a$ and $\alpha_b$ ($\alpha_b \oplus 1$) as their raw key bits if Charlie announced a detection related to a $0$ ($\pi$) phase difference between Alice's and Bob's phases.

This holds for protocols in~\cite{LYDS18,Ma.,Tamaki.} with minor differences. For the protocol in \cite{Wang.2018}, it holds in the $X$ basis. The raw key bit, however, is obtained from the $Z$ basis when single clicks are announced by Charlie, irrespective of which detector clicked.
For the protocols in \cite{CYW+18,CAL18}, it holds with $\Delta=0$ in the encoding basis.

\smallskip

\textit{Parameter Estimation} -- A raw key is formed by concatenating the raw key bits obtained in the previous step. All the remaining data unrelated to the key bits can be fully disclosed to estimate the decoy-state parameters related to security.

Up to minor differences, this step is the same in all TF-QKD protocols.
In fact, they all use the decoy-state technique~\cite{Hwang.2003,Wan05,LMC05} to estimate the single-photon quantities related to security, with the exception of \cite{LL18}, which was already kept out of our description.
Even in~\cite{Ma.} decoy states are extensively used to estimate the photon-number dependent quantities appearing in the phase error rate of the protocol. In this case, having only three intensities for $\mu_{a,b}$ might be insufficient to obtain a tight estimation of the phase error rate.
A similar argument applies to ref.~\cite{CYW+18}, where four intensity levels were used to obtain a good key rate.
Here we find that three intensity levels are sufficient to extract good key rates from the protocols in Refs.~\cite{LYDS18,Wang.2018} and \cite{CAL18}.

\smallskip

\textit{Key distillation} -- The users run classical post-processing procedures such as error correction and privacy amplification to distil the final secure key from the raw key.

The amount of privacy amplification in this step is specific to each TF-QKD protocol as it depends on the detailed security analysis. In the present work, we only consider two specific key distillation rates for exemplificative purposes, given in Refs.~\cite{LYDS18,Wang.2018}. For both protocols we consider the standard decoy state equations in the asymptotic scenario, providing
the lower bound for the 0-photon yield, $\underline{y}_0 = (v Q_w e^w - w Q_v e^v)/(v-w)$;
the lower bound for the 1-photon yield, $\underline{y}_1 = [u^2 Q_v e^v - u^2 Q_w e^w - (v^2-w^2) ( Q_u e^u - \underline{y}_0 )] / [u(u v-u w-v^2+w^2)]$;
and the upper bound for the 1-photon error rate, $\overline{e}_1 = (E_v Q_v e^v - E_w Q_w e^w)/[(v-w)\underline{y}_1]$.
In these equations, $u=u_a+u_b$, $v=v_a+v_b$ and $w=w_a+w_b$ are the total intensity values for the signal state, the decoy state and the vacuum state, respectively.
For the total intensity, we also use the symbol $\mu=\mu_a+\mu_b=\{u,v,w\}$.
In our experiment, we set $u=0.4$, $v=0.16$ and $w=10^{-5}$ for the protocols in \cite{LYDS18,Wang.2018} and $u_a=u_b=0.02$, $v_a=v_b=0.2$, $w=10^{-5}$ for the protocol in \cite{CAL18}.
The parameters $\underline{y}_1$ ($\underline{y}_0$) and $\overline{e}_1$ are, respectively, the lower bound for the single-photon (zero-photon) yield and the upper bound for the single-photon phase error rate; $Q_\mu$ and $E_\mu$ are the gain and QBER measured by detector D1 in Fig.~1
a.
Then for the original TF-QKD key rate we use
\begin{equation}\label{SKR_TFQKD}
 R^{\prime}=\{\underline{Q}_1 [1-h(\overline{e}_1)] - f_{EC} Q_{u} h(E_u) \}/M^{\prime}.
\end{equation}
Here, $\underline{Q}_1 = \mu e^{-\mu} \underline{y}_1$ is the lower bound for the single-photon gain;
$f_{EC}$ is the error correction factor, set equal to 1.15 in our simulations;
$h$ the binary entropy function;
$M^{\prime}=M/2$ and $M$ is the number of phase slices used to reconcile the global random phase, set equal to 16 in our simulations.
In Eq.~\eqref{SKR_TFQKD}, $E_u$ includes $E_M$, the intrinsic misalignment error of TF-QKD, which is equal to 1.275\% for $M=16$~\cite{LYDS18}.
The key rate in Eq.~\eqref{SKR_TFQKD} is secure under the conditions clarified in ref.~\cite{LYDS18}. When these conditions are not met, Eve can perform other attacks like the `collective beam splitting' (CBS) attack~\cite{LYDS18} or the one described in \cite{WHY18}.

For the `Send-Not Send' TF-QKD protocol by Wang \textit{et al.}~\cite{Wang.2018} we use the key rate equation
\begin{equation}\label{SKR_SNSTFQKD}
 R^{\prime\prime}=\underline{Q}_0^z + \underline{Q}_1^z [1-h(\overline{e}_1)] - f_{EC} Q^z h(E^z),
\end{equation}
with $Q^z = \epsilon^2 Q_{u} + \epsilon (1 - \epsilon) (Q_{u_a} + Q_{u_b}) + (1 - \epsilon)^2 Q_0$, $ E^z  = [\epsilon^2 Q_{u} + (1 - \epsilon)^2 Q_0]/{Q^z}$.
The parameter $\epsilon$ has been defined in the state preparation step of the protocol; $Q_{u}$, $Q_{u_a}$, $Q_{u_b}$, $Q_{0}$ are the gains (i.e. ratio of successfully detected events to sent optical pulses) of the protocol, measured in the experiment when both of the users, only Alice, only Bob, none of the users, respectively, send out optical pulses.
The values for these quantities at each attenuation level are reported the Supplementary Information.
The single-photon gain in the $Z$ basis assumes the form $\underline{Q}_1^z = \[\epsilon (1 - \epsilon) (u_a e^{-u_a}+u_b e^{-u_b}) + \epsilon^2 (u e^{-u})\] \underline{y}_1$ and $\underline{Q}_0=\[ (1-\epsilon)^2 + \epsilon (1-\epsilon) (e^{-u_a}+e^{-u_b}) + \epsilon^2 e^{-u} \]\underline{y}_0$.
The single-photon quantities $\underline{y}_0$, $\underline{y}_1$ and $\overline{e}_1$ are drawn from the $X$ basis of the protocol using the equations written above.
In this specific protocol, the number of phase slices is large, leading to no misalignment error in the $X$ basis.
Also, unlike the original send-not send TF-QKD~\cite{Wang.2018}, we consider here three intensities to implement the decoy-state technique, which is practical.
Finally, we include an extra term $\underline{Q}_0^z$ in the key rate equation \eqref{SKR_SNSTFQKD}, accounting for the fact that Eve cannot extract any useful information from the vacuum pulses prepared by the users~\cite{Lo05,Koa06}.

For the protocol by Curty \textit{et al.}~\cite{CAL18} we use the key rate equation written for their Protocol~3, after adding the error correction factor $f_{EC}$, to make the proposal more practical:
\begin{equation}\label{SKR_Curty}
  R^{\prime\prime\prime} = Q^z [1-h(\overline{e}^x_{1})] - f_{EC} Q^z h(E^z).
\end{equation}
This extra term makes the key rate smaller, so it is even more difficult to overcome the SKC$_0$ when it is taken into account.
The counts for the raw key come from detector D1 in the setup of Fig.~1
a, as for the other protocols.
Most quantities in Eq.~\eqref{SKR_Curty} are similar to those already introduced for the other protocols,  with the exception of the phase error rate $\overline{e}^x_{1}$, which deserves a specific discussion.
It has been taken from Eq.~(15) of ref.~\cite{CAL18} and amounts to
\begin{align}\label{e1Cur}
\nonumber  \overline{e}^x_{1} &=\frac{1}{Q^z}\sum_{j=0,1}\[\sum_{m,n=0}^{\infty} c_{m}^{(j)} c_{n}^{(j)}\sqrt{\overline{Y}^x_{mn}}\]^2 \\
\nonumber  &\leq \frac{1}{Q^z}\sum_{j=0,1}\[\sum_{m,n=0}^{\infty} c_{m}^{(j)} c_{n}^{(j)}\sqrt{g_{mn}(\overline{Y}^x_{mn},Y_{\textrm{cut}})}\]^2\\
  &\simeq \frac{1}{Q^z}\sum_{j=0,1}\[\sum_{m,n=0}^{N_{\textrm{cut}}} c_{m}^{(j)} c_{n}^{(j)}\sqrt{g_{mn}(\overline{Y}^x_{mn},Y_{\textrm{cut}})}\]^2.
\end{align}
In Eq.~\eqref{e1Cur}, the coefficient $c_{k}^{(0)}$ ($c_{k}^{(1)}$) is defined as $c_{k}^{(0)}=e^{-\mu/2} \mu^{k/2}/\sqrt{k!}$ when the integer $k$ is even (odd) and $0$ otherwise~\cite{CAL18};
$g_{mn}(\overline{Y}^x_{mn},Y_{\textrm{cut}})$ is a function equal to $\overline{Y}^x_{mn}$ if $m+n<Y_{\textrm{cut}}$ and equal to $1$ otherwise;
$Y_{\textrm{cut}},N_{\textrm{cut}}$ are two integers such that $Y_{\textrm{cut}}<N_{\textrm{cut}}$.
In our experiment we set $Y_{\textrm{cut}}=5$ and $N_{\textrm{cut}}=20$.
The quantities $\overline{Y}^x_{mn}$ are upper bounds for the yields obtained when Alice sent $m$ photons and Bob sent $n$ photons.
These are estimated using a constrained optimisation linear program within the standard decoy state technique~\cite{Wan05,LMC05}, with the only difference that the yields have to be maximised rather than minimised in order to provide the worst-case phase error rate.
For the quantity $\overline{Y}^x_{01}$, for example, we look for the maximum of $Y^x_{01}$ in the interval $[0,1]$ constrained by the gains $Q^x_{\mu_A,\mu_B}$ that are measured when Alice prepares the intensity $\mu_A$ and Bob prepares the intensity $\mu_B$ in the decoy basis $X$.
The intensities prepared by the users take on three values: $u$ (signal), $v$ (decoy) and $w$ (vacuum).
A positive and nearly optimal key rate (i.e. close to asymptotic scenario) is found when we set $u=0.02$, $v=0.2$ and $w<10^{-5}$, i.e., when the intensity of the decoy state is larger than the one of the signal state.
Moreover, we measured all the intensity combinations $uu$, $uv$, $uw$, $vv$, $vw$ and $ww$ to improve the decoy-state estimation.
This method with only three intensities was not disclosed in the original protocol~\cite{CAL18} and we propose it here for the first time.
The simulations agree well with the experimental data, providing an SKR of 271.3~bit/s.
}

\begin{acknowledgments}
We acknowledge useful discussions with Marcos Curty about the protocol in ref.~\cite{CAL18}.
 M.M. acknowledges financial support from the Engineering and Physical Sciences Research Council (EPSRC) and Toshiba Research Europe Ltd.
 M.P. acknowledges funding from the European Union's Horizon 2020 research and innovation programme under the Marie Sk\l{}odowska-Curie grant agreement No 675662.
 G.L.R. acknowledges financial support via the EPSRC funded Centre for Doctoral Training (CDT) in Integrated Photonic and Electronic Systems, Toshiba Research Europe Ltd. and The Royal Commission for the Exhibition of 1851.
\end{acknowledgments}

\bibliography{TF-QKD_Bibliography}

\clearpage
\onecolumngrid
\section*{Supplementary Information for\\Experimental quantum key distribution beyond the repeaterless secret key capacity}

\author{M. Minder$^{1,2 \dagger}$, M. Pittaluga$^{1,3 \dagger}$, G. L. Roberts$^{1,2}$, M. Lucamarini$^{1\star}$, J. F. Dynes$^{1}$, Z. L. Yuan$^{1}$ \& A. J. Shields$^{1}$}

\begin{center}

{M. Minder$^{1,2 \dagger}$, M. Pittaluga$^{1,3 \dagger}$, G. L. Roberts$^{1,2}$, M. Lucamarini$^{1\star}$, J. F. Dynes$^{1}$, Z. L. Yuan$^{1}$ \& A. J. Shields$^{1}$}\\

\smallskip

\textit{$^{1}$Toshiba Research Europe Ltd, 208 Cambridge Science Park, Cambridge CB4 0GZ, UK \\
$^{2}$Department of Engineering, Cambridge University, 9JJ Thomson Avenue, Cambridge CB3 0FA, UK \\
$^{3}$School of Electronic and Electrical Engineering, University of Leeds, Leeds LS2 9JT, UK \\
$^{\dagger}$These authors contributed equally to this work. \\
$^{\ast}$email: marco.lucamarini@crl.toshiba.co.uk\\}
\end{center}

\bigskip

\twocolumngrid

\section*{Optical Phase-Locked Loop}

\begin{figure}[h!]
    \centering
    \includegraphics[width=0.75\columnwidth]{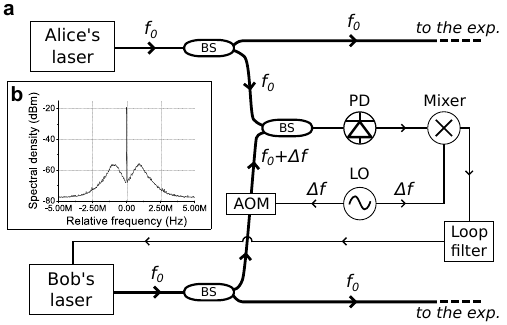}
    \caption{
    \textbf{Schematics of the heterodyne OPLL to lock Bob's laser to Alice's. a,}
    BS, 50/50 beam splitter;
    PD, photodiode;
    Mixer, electronic mixer used as phase detector;
    LO, electronic local oscillator;
    $f_0$, reference frequency;
    $\Delta f$, frequency offset (80~MHz) provided by the LO.
    \textbf{b,} RF output of the photodiode as recorded by a spectrum analyser when the two lasers are phase-locked.
    An 80~MHz offset is applied to the graph horizontal axis.
    The measurement shows a $\sim$~40~dB extinction ratio leading to a residual phase error of $\sigma^2_\phi = 7.53 \cdot 10^{-3}$ rad$^2$.
    }
    \label{fig:OPLL_schematic}
\end{figure}

In the setup of Fig.~1a, Bob's laser (slave) is locked to Alice's (master) through a heterodyne optical phase-locked loop (OPLL), which we reproduce in Fig.~\ref{fig:OPLL_schematic} in more detail.
This is a well-established technique~\cite{Bordonalli.1999,Satyan.2009} to lock two lasers with a fixed frequency offset.
The light beams from master and slave interfere on a beam splitter (BS), generating a beating signal that is recorded by a photodiode (PD) and compared with a fixed frequency from an electrical local oscillator (LO), operated at a frequency offset between the two lasers $\Delta f=80$~MHz.
This offset is imparted to the light coming from Bob's laser through an acousto-optic modulator (AOM).
A phase detector measures the phase shift between the beating signal and the LO and produces an error signal that is fed into a loop filter.
The loop filter processes this signal and tunes the emission frequency of the slave laser so as to minimise it.
When the two lasers are successfully locked with each other, the slave laser follows the free-running master's fluctuations, so that the frequency offset between the two remains constant.

Heterodyne OPLL has been chosen over its homodyne analogue to have the beat note recorded from the PD far from DC and close to the frequency of the LO.
This allows us to filter out the low-frequency noise and to make the locking mechanism more robust against intensity fluctuations.
In the current experimental implementation, the residual phase noise associated with the OPLL is $\sigma^2_\phi = 7.53 \cdot 10^{-3}$ rad$^2$ (see Fig.~\ref{fig:OPLL_schematic}b), equivalent to a phase error of $4.97^{\circ}$.

\section*{Bit encoding and phase randomisation}

\begin{figure}[b]
    \centering
    \includegraphics[width=0.7\columnwidth]{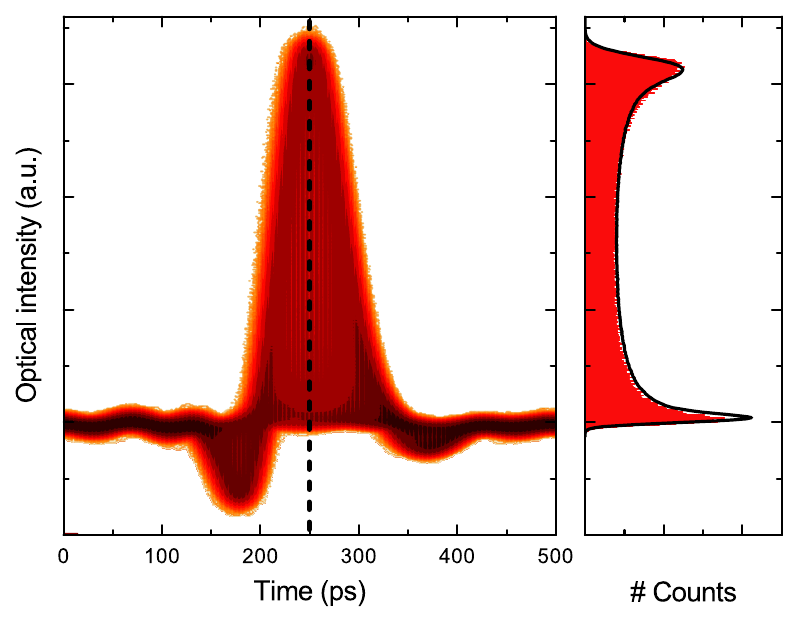}
    \caption{
    \textbf{Phase randomisation -- Support experiment.}
    \textit{Left:} colour coded density plot of Alice's slave laser intensity after a 500-ps AMZI, when injected by a gain switched master laser.
    \textit{Right:} histogram of the optical intensity, recorded along the dashed line in the left figure.
    Also shown is the simulation line that accounts for experimental imperfections (solid black line).
	The agreement between the experimental results and simulation indicates that the pulses have random phase~\cite{Roberts.2017b}.
	An analogous measurement on Bob's slave laser gives similar results.
    }
    \label{fig:Injection_Locking_phase_randomisation}
\end{figure}

There are two kinds of information to be phase encoded: the bit/basis and the global random phase.
For that, we use two phase modulators (PMs), one for each user, driven by a 1~GHz square wave.
The amplitude of the driving signal is related to the phase encoded onto the optical pulses.
The PMs are driven using two channels of 8-bit DACs, synchronised with the lasers.
For the bit/basis encoding, we test 4 different values corresponding to $\{0, \pi/2, \pi,3\pi/2\}$.
For the global random phase, we add a pseudo-random phase on top of the bit-encoded pulses.
The random phase comes from a $2^{10}$-long string of values randomly chosen among $2^5$ evenly spaced values of the $[0,2\pi)$ interval.

Alongside the main experiment, a different experiment was performed to support its findings.
In the support experiment, the users' lasers were phase-locked to a common master laser by means of optical injection locking (OIL)~\cite{YPJ+03,CLF+16a}.
The master laser was gain-switched to generate pulses that were fully phase-randomised~\cite{JCS+11,YLD+14}.
The random phase was passed onto the slave lasers by the OIL mechanism.
In the support experiment, we first verified the effectiveness of the phase randomisation by interfering pulses of adjacent clock cycles onto a 500-ps asymmetric Mach-Zehnder interferometer (AMZI).
This produced the histogram in Fig.~\ref{fig:Injection_Locking_phase_randomisation}, which closely resembles the one expected from phase-randomised laser fields~\cite{JCS+11,YLD+14,Roberts.2017b}.
We then confirmed the phase coordination between the users by interfering pulse pairs emitted by the users' lasers at the same clock and obtained a first-order interference visibility of 98.7\%, showing that active discrete phase randomisation and passive continuous phase randomisation generate a similar visibility.
Both the main and the support experiments were run at 2~GHz with 50\% of the optical pulses acting as unmodulated phase reference pulses to stabilise the setup.

\section*{Phase stabilisation feedback and QBER model}
\begin{figure}[b]
    \centering
    \includegraphics[width=0.75\columnwidth]{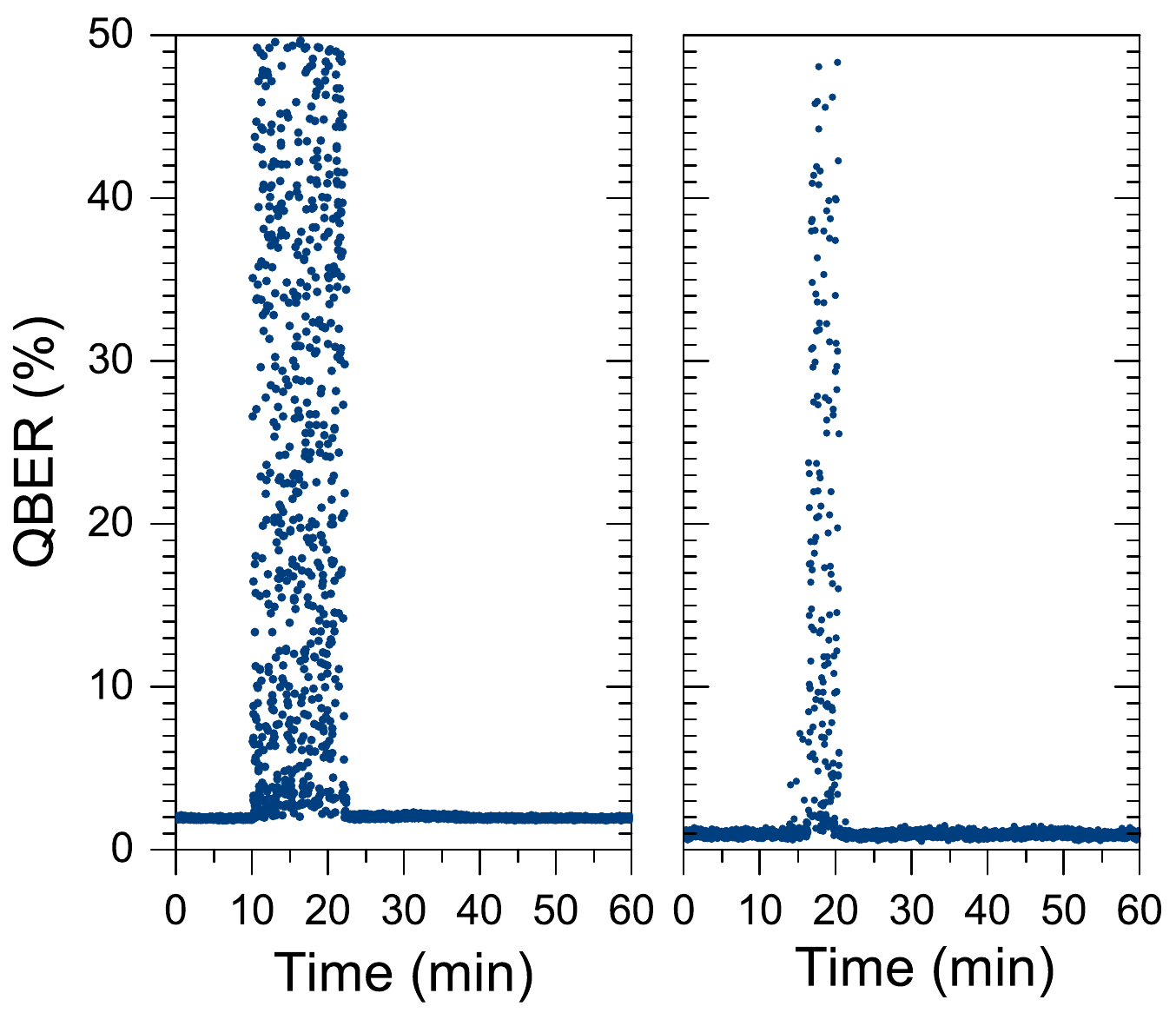}
    \caption{
    \textbf{Phase stabilisation.}
    For both the main (\textit{left}) and the support (\textit{right}) experiment, the phase feedback system was turned OFF and ON again.
    Without the phase feedback, the relative phase of the optical fields drifts freely, causing QBER fluctuations between 0 and 50\%. Due to the bit-flip symmetry of the QBER, we show its complementary value $1-$QBER whenever it overcomes 50\%.
    }
    \label{fig:phase_feedback}
\end{figure}

\begin{figure}
    \centering
    \includegraphics[width=0.85\columnwidth]{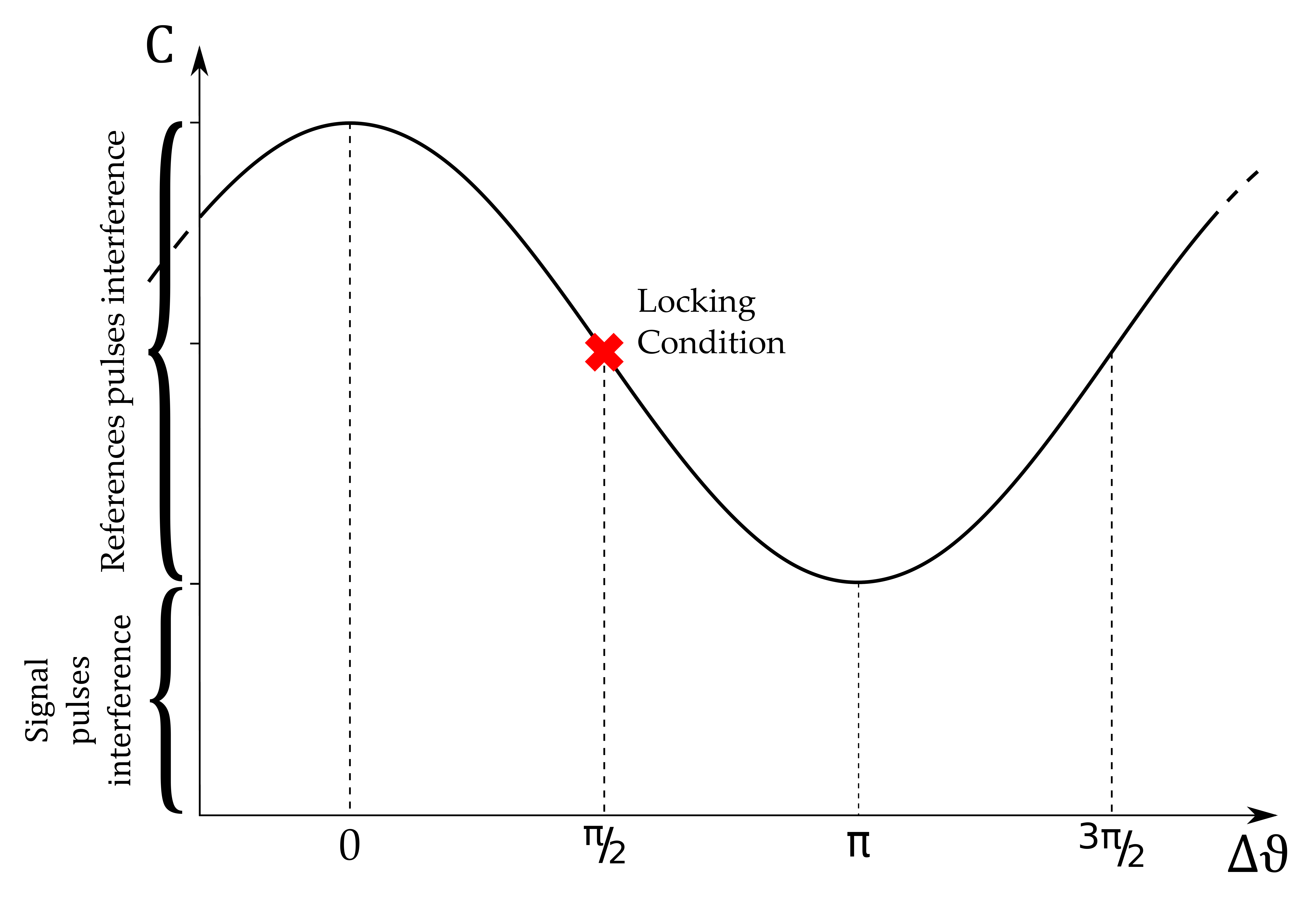}
    \caption{
    \textbf{Feedback locking condition.}
    Schematic of the number of counts recorded at D3 as function of the phase offset between the two quantum channels.
    A red cross marks the $\pi/2$ phase locking condition.
    }
    \label{fig:Feedback_Interference}
\end{figure}

The effectiveness of the control system for both the main and the support experiment is illustrated in Fig.~\ref{fig:phase_feedback}.
For each experiment we plot the QBER of the system as a function of time, for a channel loss of 30.1~dB between Alice and Bob.
For illustration purposes, the feedback was first enabled for about 10-15 minutes in both cases.
During this period, the measured QBER was kept at an average value of 1.8\% and 1\% in the main and support experiments, respectively.
Then, the feedback was disabled for few minutes, causing large fluctuations. Re-enabling the feedback restored the low QBER.

\begin{figure}[tpb]
    \centering
    \includegraphics[width=0.85\columnwidth]{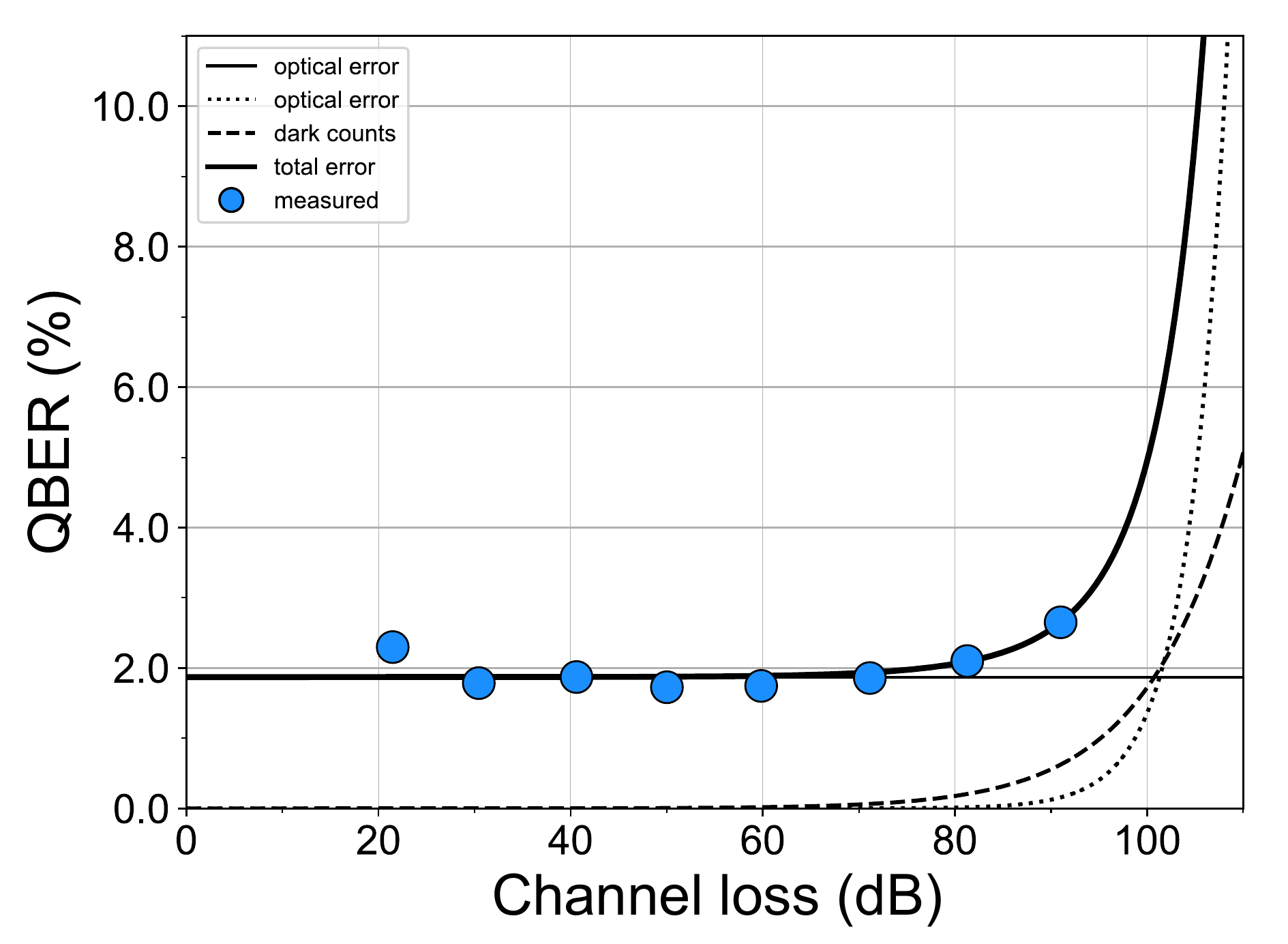}
    \caption{
    \textbf{QBER model.}
    QBER retrieved experimentally (symbols) alongside their theoretical simulation (solid line).
    Also plotted are QBER contributions from the optical error (thin solid line), the detector dark count (dashed line) and the feedback error (dotted line).
    }
    \label{fig:QBER_Contributes}
\end{figure}

To stabilise the setup, we used the count rate of D3 (see Fig.~1a) as feedback signal.
With short integration intervals, the counts detected by D3 can be written as
\begin{equation}\label{eq:Interference_eq}
  C = C_0 + C_1 (1-\cos \Delta \vartheta),
\end{equation}
where $C_0$ represents the count floor, contributed from the interference of the encoded pulses in addition to the detector dark counts, while $C_1$ is the amplitude of the interference between the unmodulated pulses, and $\Delta \vartheta$ is the phase difference which drifts rapidly due to the fibre length variations.
Here, $C_0 \approxeq  C_1$ as we set equal intensity and probability between the encoded and reference pulses in the experiment.
The locking point for the stabilisation is chosen at the quadrature point ($\Delta \theta = \pi/2$), as shown in Fig.~\ref{fig:Feedback_Interference}.

This is achieved by varying the DC bias of the PM (Fig.~1b), such that the phase drift can be counteracted and a constant count rate of $C_0$ + $C_1$ can be maintained.
The phase compensation is provided by a PID controller that receives as input the number of photons collected by the photon counter connected to detector D3 and that modifies the PM's DC offset through its amplified 12-bit DAC.
The DC offset is corrected every 10-100 ms, with the precise correction rate optimised for each attenuation value to minimise the QBER fluctuations.
Near the locking point, the count rate is approximately a linear function of the phase drift and therefore allows precise phase compensation at the high count rate limit.
However, the feedback count decreases exponentially with the Alice-Bob channel loss ($\tilde{L}$ in dB) as $e^{-\tilde{L}/20}$ meaning that the shot noise in the count rate $C$ will become increasingly non-negligible.
This introduces an estimation error in the phase drift that is on average
\begin{equation}\label{eq:Feedback_phase_error}
  \Delta \vartheta_e \approx 2/\sqrt{C}
\end{equation}
and still allows for an approximate phase correction at high loss.

We model the experimentally measured QBER with the results shown in Fig.~\ref{fig:QBER_Contributes}.
To simulate the feedback error,  we assume that the phase estimation error causes a persistent phase misalignment that is proportional to $\Delta\vartheta_e$, with a coefficient as the only fitting parameter used in the entire model.
All other parameters were directly taken from the experiment.
Besides the feedback error, we take into account two more error sources.
The optical error, arising from the finite interference visibility and any modulation error in the electrical signals, contributes to a baseline QBER (thin solid line).
The detector dark count (dashed line) contributes to the increase in the QBER starting from channel losses greater than 60~dB.
The feedback error affects mainly the last point, at approximately 90~dB attenuation.
The QBER model includes all the sources of error and well reproduces the experimental data.

\onecolumngrid
\section*{Detailed experimental results}

\makeatletter
\def\hlinewd#1{%
  \noalign{\ifnum0=`}\fi\hrule \@height #1 \futurelet
   \reserved@a\@xhline}
\makeatother

\begin{table}[!ht]
\centering
\setlength{\tabcolsep}{0mm}
\begin{tabular}{cc|cc|ccc|c|c|c}

\hline\hline
\multicolumn{2}{c|}{Alice $\rightarrow$ Charlie} & \multicolumn{2}{c|}{Bob $\rightarrow$ Charlie} & \multicolumn{3}{c}{Alice $\&$ Bob $\rightarrow$ Charlie} & \multicolumn{2}{|c}{~~~~~~~Secret key rate}\\
\hline
\multicolumn{1}{c}{Attenuation} & \multicolumn{1}{c|}{Gain} & \multicolumn{1}{c}{Attenuation} & \multicolumn{1}{c|}{Gain} & \multicolumn{1}{c}{Attenuation}
& \multicolumn{1}{c}{Gain} & \multicolumn{1}{c|}{QBER} & \multicolumn{1}{c}{ref.~\cite{LYDS18}} & \multicolumn{1}{c}{~ref.~\cite{Wang.2018}~} & \multicolumn{1}{c}{~SKC$_0$~}\\

\multicolumn{1}{c}{(dB)} & \multicolumn{1}{c|}{$(\times10^{-6})$} & \multicolumn{1}{c}{(dB)} & \multicolumn{1}{c|}{$(\times10^{-6})$} & \multicolumn{1}{c}{(dB)} & \multicolumn{1}{c}{$(\times10^{-6})$} & \multicolumn{1}{c|}{\%} & \multicolumn{2}{c}{$~~~~~~(\times10^{3})$ bit/sec}\\

\hline\hline

10.7 & 3000.8 & 10.8 & 3149.9 & 21.5 & 5562.8 & 2.29 & 159 & 74.5 & ~10,250~ \\
\rowcolor{LightGray}
     & 1154.4 &      & 1228.7 &      & 2172.4 & 2.20 & & & \\

15.3 & 993.4  & 15.2 & 985.8  & 30.5 & 1984.0 & 1.79 & 66.1 & 30.4  & 1,286 \\
\rowcolor{LightGray}
     &  382.2 &      &  388.3 &      &  765.7 & 1.96 & & & \\

20.4 &  301.1 & 20.3 &  293.8 & 40.7 &  592.1 & 1.87 & 19.2 & 8.86  & 122.8 \\
\rowcolor{LightGray}
     &  117.2 &      &  117.0 &      &  233.4 & 1.99 & & & \\

25.1 &   95.4 & 25.0 &  100.4 & 50.1 &  195.6 & 1.73 & 6.71 & 3.03  & 14.10 \\
\rowcolor{LightGray}
     &   37.2 &      &   39.3 &      &  76.2 & 1.83 & & & \\

30.0 &   30.2 & 29.9 &   63.4 & 61.7 &   61.6 & 1.75 & 2.14 & 0.933  & 1.476 \\
\rowcolor{LightGray}
     &   12.1 &      &   12.0 &      &   24.6 & 1.71 & & & \\

35.6 &   8.74 & 35.5 &   9.44 & 71.1 &   18.2 & 1.86 & 0.602 & 0.213  & 0.112 \\
\rowcolor{LightGray}
    &    3.48 &      &   3.72 &      &   7.19 & 1.97 & & & \\

40.6 &   2.84 & 40.6 &   2.86 & 81.2 &   5.65 & 2.10 & 0.163 & 0.0176  & 0.011 \\
\rowcolor{LightGray}
     &   1.14 &      &   1.20 &      &   2.32 & 2.40 & & & \\

45.4 &   0.91 & 45.4 &   0.91 & 90.8 &   1.79 & 2.65 & 0.045 & -  & 0.001 \\
\rowcolor{LightGray}
     &  0.353 &      &   0.368 &      &  0.72 & 3.57 & & & \\
\hline\hline
\end{tabular}
\caption{Numerical data for the main experiment with the TF-QKD protocols in refs.~\cite{LYDS18,Wang.2018}.
The white (grey) rows in the first, second and third column report the values for the signal gains $Q_{u_a}$, $Q_{u_b}$, $Q_{u}$ (decoy gains $Q_{v_a}$, $Q_{v_b}$, $Q_{v}$), respectively, registered by detector D1 in Fig.~1a when only Alice, only Bob or both users send pulses to the intermediate node.
When no user sends out pulses, the measured gain is $Q_0=25.9\times10^{-9}$.
The flux set by each user was $u_a=u_b=0.2$ photons per pulse for the signal states and $v_a=v_b=0.08$ photons per pulse for the decoy states.
The total vacuum is set to $w=10^{-5}$.}
\label{table:gainandQBER_new}
\end{table}

\begin{table}[!ht]
\centering

\begin{tabular}{c|cc}

\hline\hline

\multicolumn{3}{c}{Phase randomised}\\
\hline
\multicolumn{1}{c|}{Flux} & \multicolumn{2}{c}{Gain (1 detector)} \\
\multicolumn{1}{c|}{ } & \multicolumn{2}{c}{$(\times10^{-6})$} \\
\hline\hline

uu & \multicolumn{2}{c}{1.71} \\
vv & \multicolumn{2}{c}{18.2} \\
ww & \multicolumn{2}{c}{0.026} \\
uv & \multicolumn{2}{c}{8.77} \\
uw & \multicolumn{2}{c}{0.913} \\
vw & \multicolumn{2}{c}{8.74} \\

\hline\hline

\multicolumn{3}{c}{Encoded}\\
\hline
Flux & Gain & QBER \\
 & $(\times10^{-6})$ & \% \\
\hline\hline

uu & 1.79 & 2.65 \\

\hline\hline

\end{tabular}
\caption{Measured quantities for the protocol in ref.~\cite{CAL18}. At 71.1~dB channel loss it provides a SKR of $270.7$~bit/s, which is 2.42 times above the ideal SKC$_0$ bound at the same attenuation ($112.0$~bit/s).
The flux set by each user was $u_a=u_b=0.02$ photons per pulse for the signal states and $v_a=v_b=0.2$ photons per pulse for the decoy states. The total vacuum was set to $w=10^{-5}$.}
\label{table:Curty_data}
\end{table}

\end{document}